\documentstyle[preprint,aps]{revtex}
\begin{document}
\draft
\title{Phase separation of edge states in the integer quantum Hall regime}
\author{L. Brey, J. J. Palacios, and C. Tejedor}
\address{Departamento de F\'{\i}sica de la Materia Condensada.
Universidad Aut\'onoma de Madrid.
Cantoblanco, 28049, Madrid. Spain. }
\maketitle
\begin{abstract}
Coulomb effects on the edge states of a two dimensional electron gas in
the presence of a high magnetic field are studied for different
widths of the boundaries. Schr\"odinger and Poisson equations
are selfconsistently solved in the integer Quantum Hall regime.
Regions of flat bands at the Fermi level appear for smooth
interfaces in order to minimize the electrostatic energy related to the
existence of dipoles induced by the magnetic field. These plateaus
determine the phase separation in stripes of compressible and
incompressible electron liquids.
\end{abstract}

A high magnetic field perpendicular to a two dimensional (2D) electron
gas produces a spectrum with a continuous part in which states located
close to the boundaries are involved.
Those edge states are essential for the response of the
system to external perturbations, particularly in carrying a net
electrical current when a bias is applied to the sample\cite{1}. For an
empty system (or when electron-electron interaction is neglected),
edge states are completely determined by the confining potential.
However, when a density of electrons populates the system, Coulomb
interaction changes qualitatively the band structure. Such a problem
has been only addressed in the case of a very smooth external potential
either by a Thomas-Fermi approach\cite{2,3} or by solving the Poisson
equation in the classical limit\cite{4}. Spin splitting effects have
also been studied in a variational scheme\cite{5}. In some cases, like
in the case of the Fractional Quantum Hall regime, in which many-body
effects are crucial, only general arguments have been given to
discuss the importance and behavior of edge states\cite{6,7,8,9,10}.
In all those works, the smoothness of the confining potential is a
necessity due to the strong simplifications involved. The actual
situations in experiments cover the whole range from abrupt
(etching techniques) to smooth (gate techniques)
boundaries\cite{1}. To understand the properties of the edge states for
any width of the interface region a quantum analysis including
electrostatic effects is needed. This implies to solve selfconsistently
the Schr\"{o}dinger and Poisson equations for electrons confined
by a potential defined in a 2D system in the presence
of the external magnetic field. \\
\par We are interested in studying the properties of electrons confined
in the $z$-direction by semiconductor interfaces or quantum wells
affected by a magnetic field $\vec{B}=B \vec{u _ z }$. The system has no
constrictions in the $y$-direction while in the $x$-direction there are
some boundaries with typical widths significantly larger than the
extent of wave functions in the $z$-direction. Therefore, we neglect
the effects of the width in $z$ of the electron gas and just consider a
strictly 2D ($xy$) system with boundaries in the $x$-direction.
Physically, such boundaries are produced by gate potentials which
deplete the 2D gas in some regions confining the electrons in the rest
of the $xy$ plane. We have a total depletion under the gates by putting
infinite barriers in $x=0$ and $x=W$, while the smoother interface
region is the sum of two terms: the electrostatic potential produced by
the electronic charge, and a confining potential in the $x$-direction
which we take as the one created by a fictitious distribution of
positive charge with the trapezoidal shape
\begin{eqnarray}
\rho _ {+} (x)= \frac {n _ {+}}{L} \left\{ \begin{array}{ll}
x & \; \; \; \; 0<x<L \\ L & \; \; \; \; L<x<W-L \\ W-x & \; \; \; \;
W-L<x<W \end{array} \right.
\end{eqnarray}
In this way, the electrons with a 2D density $n _ +$
are smoothly confined in a channel of width $W$. With this model, the
Landau gauge $\vec{A} =(0,Bx,0)$ is the adequate to study the problem.
The wave functions have the form
\begin{eqnarray}
\Psi _ {n,k} (x,y)= \frac {e^{iky}}{ \sqrt {L _y}} \phi ^\sigma _{n,k}
(x) \end{eqnarray}
where the wave function $ \phi ^\sigma _{n,k} (x)$ is an eigenstate,
with eigenvalue $\varepsilon ^\sigma _{n,k} $, of the one-dimensional
Schr\"{o}dinger equation \begin{eqnarray}
\left[ - \frac {\hbar ^2}{2m^\ast } \frac {\partial ^2}{\partial x^2}+
\frac {m^ \ast \omega _c ^2 (x-x_0)^2}{2}+eV(x)+ \frac { \sigma
g^\ast \mu _B B}{2} \right] \phi ^\sigma _{n,k} (x)= \varepsilon
^\sigma _ {n,k} \phi ^\sigma _{n,k} (x)
\end{eqnarray}
where $\omega _c =eB/m ^\ast $ is the cyclotron frequency, $x_0=k
\ell ^2$ is the semiclassical center of the orbit in terms of the
magnetic length $\ell = \sqrt {( \hbar /eB)}$, $\sigma $ is the spin
of the electron, $g^\ast$ is the effective $g$-factor and $\mu _B$ the
Bohr magneton. The selfconsistent electrostatic potential is given by
\begin{eqnarray}
V(x) = \frac {2e}{\epsilon}
\int dx' [ \rho _+ (x') - \rho (x') ] \ln |x-x'|
\end{eqnarray}
$\epsilon $ being the dielectric constant of the medium and
the electronic selfconsistent charge density being given by
\begin{equation}
\rho (x)= \sum _{ \sigma ,n} \int ^{E_F} d \varepsilon ^\sigma _ {n,k}
| \phi ^\sigma _ {n,k} (x)| ^2
\end{equation}
$E_F$ is the Fermi level which, for given total charge density, must be
computed within the iterative procedure to solve the Schr\"{o}dinger
equation. We define the total charge by means of the filling
factor $\nu $ for the 2D bulk (i.e. the centre region of the wide
channel). The iterations are started by taking the electronic charge
locally equal to the positive background, i.e. with a potential $V(x)$
which is flat in between the two infinite barriers. We neglect the
exchange interaction between electrons, therefore we restrict our
calculation to even filling factors where exchange effects are less
important\cite{11}. Electrostatic effects are the same for any
$\nu $ corresponding to an incompressible liquid in bulk, i.e. with a
large enough gap at the Fermi level. Therefore, it is
possible to extend our conclusions to odd filling factors. \\
\par Since it is very interesting to discuss what happens to the
different branches of the dispersion relation, we present
results for the case $ \nu =6 $ and $B=1T$. We use typical parameters
for GaAs so that $g^\ast $ is so small that there is no difference
between spin-up and spin-down bands. In order to analyze the
importance of the boundary smoothness figures 1 to 3 give the dispersion
relations $\varepsilon _{n,k} \equiv \varepsilon _n(x_0)$, potential
profiles $V(x)$ and charge densities $\rho (x)$ and $\rho _+ (x)- \rho
(x)$ (multiplied by 50 ) for the cases $L=10 \ell $, $L=30 \ell $
and $L=50 \ell $ respectively. All the figures give results computed
with $W=2L+10 \ell $ in which the two interfaces are decoupled from each
other as we have checked out by increasing the separation (changing the
factor $10$ by higher numbers) and obtaining the same results. Due to
the symmetry with respect
to the centre of the channel we only present results for $x<W/2$.
{}From figure 1, one observes that an interface of ten magnetic
lengths can be considered rather abrupt because no plateaus at the edge
appear in $V(x)$ or $\varepsilon _n (x _0)$ as suggested for smooth
interfaces\cite{4}. Only some inflections appear for $L=10 \ell $ that
become flat regions for $L=30 \ell $ and still broader for the
smoother case of $L=50 \ell $. For $L=30 \ell $ and $L=50 \ell$ there
are two types of plateaus near the Fermi level which are
intrinsically different from each other. Those spatially close to the
edges are not completely occupied because they lie in energy
exactly at the Fermi level while the plateau in the bulk
is clearly below $E _ F$ so that its states are fully occupied.
Since we work in a single particle model, our results for the flat
regions at the edge present very small oscillations instead of
being completely flat so that they accommodate both electrons
and holes around the Fermi level. In our
calculations, the amplitude of such oscillations tends to zero, and
the plateaus can be considered absolutely
flat for any physical purpose, in particular for considering these
regions as locally compressible liquids corresponding to noninteger
filling factors. \\
\par Let us try to understand the physical causes for the shape of the
dispersion relation by starting with the flat regions close to the edge.
Among these plateaus, there are regions of incompressible liquid
corresponding to integer local filling factors. When $L$ increases, the
compressible regions extend while the incompressible ones remain
practically the same. This observation helps to understand the physical
origin of the spatial separation of the compressible and incompressible
phases. In a region of integer filling factor, the electronic charge
density is practically constant and the positive background varies
linearly. This implies an electrostatic dipole that increases with the
width of the incompressible region. Since that dipole means an increase
of the energy, the system reacts to minimize the energy by decreasing
the width of the incompressible region as much as possible i.e.
by forming compressible regions. In this new phase, the electronic
charge does not need to be constant any more. Therefore it can have the
same shape of the positive background (as shown in the figures) eluding
any increase of the electrostatic energy. It is important to stress that
this argument is independent of the form of a background which varies
with the position so that our results should be completely general for
any interface. In the bulk of the sample, the fictitious
positive background
is constant, the Fermi level lies above the Landau state giving
the well known incompressible liquid for integer filling factor. The
width of the incompressible phases at the interface can not decrease
up to zero because there is a lower limit imposed by the extent of the
wave functions. The minimization of the electrostatic energy,
which depends on the strength $e^2/ \epsilon \ell $ of the interaction,
dominates on any variation of kinetic energy, which depends
on the cyclotron energy $\hbar \omega _c $. \\
\par Disorder induced localization only affects to
the bulk electronic structure but not to edge states so that our results
are valid for actual samples\cite{13}.
The experiments for integer $\nu $ made on samples with boundaries
created by etching involve abrupt interfaces and the electron
liquid must be incompressible. On the contrary, for samples with
boundaries defined by gate potentials the interfaces are typically
broader than ten times the magnetic length and phase separation should
occur. The current will be carried in the compressible regions where
the zero velocity of each state compensates with the infinite density of
states. So, transport does not seem to be a good way to detect the
electrostatic effects here studied. Spectroscopic experiments are better
candidates to analyze compressible regions because the high density of
states must produce strong Fermi edge singularities in absorption and
emission of light\cite{12} as well as significant alterations in the
spectrum of edge magnetoplasmons. \\
\par This work
has been supported in part by the Comisi\'on Interministerial
de Ciencia y Tecnologia of Spain under contract MAT 91 0210 and  by the
Commission of the European Communities under contract SSC - CT 90 - 0020.

\newpage
\begin{figure}
\caption{Dispersion relations $\varepsilon _n (x_0)$ (broad continuous
lines), potential profiles
$V(x)$ (intermediate continuous line) and charge densities $\rho (x)$
(narrow continuous line) and $50[ \rho _+ (x)- \rho (x)]$ (dashed line)
as a function of the position for a boundary with $L=10 \ell $. The
Fermi level (dotted line) selfconsistently computed is also shown. }
\end{figure}
\begin{figure}
\caption{Dispersion relations $\varepsilon _n (x_0)$ (broad continuous
lines), potential profiles
$V(x)$ (intermediate continuous line) and charge densities $\rho (x)$
(narrow continuous line) and $50[ \rho _+ (x)- \rho (x)]$ (dashed line)
as a function of the position for a boundary with $L=30 \ell $. The
Fermi level (dotted line) selfconsistently computed is also shown. }
\end{figure}
\begin{figure}
\caption{Dispersion relations $\varepsilon _n (x_0)$ (broad continuous
lines), potential profiles
$V(x)$ (intermediate continuous line) and charge densities $\rho (x)$
(narrow continuous line) and $50[ \rho _+ (x)- \rho (x)]$ (dashed line)
as a function of the position for a boundary with $L=50 \ell $. The
Fermi level (dotted line) selfconsistently computed is also shown. }
\end{figure}
\end{document}